\documentclass[Physsubmission, Phys]{SciPost}

\hypersetup{
    colorlinks,
    linkcolor={red!50!black},
    citecolor={blue!50!black},
    urlcolor={blue!80!black}
}

\usepackage[bitstream-charter]{mathdesign}
\usepackage{xspace}
\usepackage{bm}
\DeclareSymbolFont{usualmathcal}{OMS}{cmsy}{m}{n}
\DeclareSymbolFontAlphabet{\mathcal}{usualmathcal}
\newcommand{\sqs}{$\sqrt{s}$}
\newcommand{\sqsn}{$\sqrt{s_{\rm NN}}$}
\newcommand{\pt}{$p_{\rm T}$\xspace}
\newcommand{\dnchdeta}{$\rm{d}\it{N}_{\rm ch}/\rm{d}\eta$\xspace}
\begin{document}

\begin{center}{\Large \textbf{
Measurements of heavy-flavor production as a function of multiplicity with ALICE at the LHC\\
}}\end{center}

\begin{center}
Yoshini Bailung\textsuperscript{1$\star$} on behalf of the ALICE Collaboration
\end{center}

\begin{center}
{\bf 1} Department of Physics, Indian Institute of Technology Indore, Madhya Pradesh, India
* yoshini.bailung@cern.ch
\end{center}

\begin{center}
\today
\end{center}


\definecolor{palegray}{gray}{0.95}
\begin{center}
\colorbox{palegray}{
  \begin{tabular}{rr}
  \begin{minipage}{0.1\textwidth}
    \includegraphics[width=30mm]{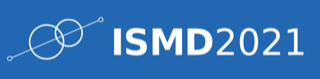}
  \end{minipage}
  &
  \begin{minipage}{0.75\textwidth}
    \begin{center}
    {\it 50th International Symposium on Multiparticle Dynamics}\\ {\it (ISMD2021)}\\
    {\it 12-16 July 2021} \\
    \doi{10.21468/SciPostPhysProc.?}\\
    \end{center}
  \end{minipage}
\end{tabular}
}
\end{center}

\section*{Abstract}
{\bf
In this contribution, the production of heavy-flavor hadrons as a function of multiplicity, via the study of the $\mathbf{D}$-meson and heavy-flavor hadron decay leptons self-normalized yields in pp collisions at the center of mass energy $\bm{\sqrt{s} = 13}$ TeV is discussed. Comparisons are made with similar measurements of J/$\bm{\psi}$ at $\bm{\sqrt{s} = 13}$ TeV and various model calculations. The $\mathbf{\Lambda_{c}^{+}/D^{0}}$ and $\mathbf{D^{+}_{s}/D^{0}}$ yield ratios in different multiplicity intervals in pp collisions at $\bm{\sqrt{s} = 13}$ TeV are also reported. In addition, the ALICE measurement of $\mathbf{\Lambda_{c}^{+}}$ production in p--Pb collisions at $\bm{\sqrt{s_{\rm{NN}}}}$ = 5.02 TeV down to transverse momentum ($\bm{p_{\rm T}}$) = 0 GeV/$\bm{c}$ is presented. Finally, the nuclear modification factor is shown for open charm hadrons at $\bm{\sqrt{s} = 5.02}$ TeV in p--Pb collisions. Finally, measurements of the elliptic flow of heavy-flavor hadron decay leptons in p--Pb systems are presented, which hint towards a possible collective behaviour in high multiplicity p--Pb collisions.
}



\section{Introduction}
\label{sec:intro}
\subsection{Physics motivation}
Heavy quarks (charm and beauty) are produced at the initial stages of the relativistic hadronic collisions via hard scattering processes. The study of heavy-flavor hadron production provides a stringent test to perturbative quantum chromodynamics (pQCD) calculations in pp collisions and can help in disentangling cold nuclear matter (CNM) effects in p--Pb collisions. Analysis of charm production as a function of charged-particle multiplicity allows the investigation of the role of multi-parton interactions (MPI), color-reconnection (CR) mechanisms, and the interplay between the hard and soft particle production in pp and p--Pb collisions. Heavy-flavor hadron production measurements in high multiplicity pp and p--Pb collisions indicate possible modification of the particle spectra compared to minimum bias measurements. These results raise questions about the collective nature of the QCD matter produced in small systems.
\subsection{Analysis Strategy}

Several detectors in ALICE~\cite{ALICE:2008ngc} are used to reconstruct charm hadrons in various rapidities ($y$) from their:
\begin{enumerate}
    \item hadronic decay channels, $\rm{D^0 \rightarrow K^- \pi^+}$; $\rm{D^+ \rightarrow K^- \pi^+ \pi^+}$; $\rm{D^{*+} \rightarrow D^0\pi^+}$; $\rm{D_{s}^{+} \rightarrow K^{-}K^{+}\pi^+}$; $\rm{\Lambda_{c}^{+}\rightarrow pK^{-}\pi^{+}}$; $\rm{\Lambda_{c}^{+}\rightarrow pK^{0}_{s}}$; $\rm{\Sigma_{c}^{0} \rightarrow \Lambda^{+}_{c}\pi^{-}}$; $\rm{\Sigma_{c}^{++} \rightarrow \Lambda^{+}_{c}\pi^{+}}$; $\rm{\Xi_{c}^{0} \rightarrow \Xi^{-}_{c}\pi^{+}}$; $\rm{\Xi_{c}^{+} \rightarrow \Xi^{-}\pi^{-}\pi^{+}}$;\\ $\rm{\Omega_{c}^{0} \rightarrow \Omega^{-}\pi^{+}}$
    \item semi-leptonic decay channels, B, D-meson $\rightarrow \rm{e} + X$; B, D-meson $\rightarrow \mu + X$; $\rm{\Xi_{c}^{0} \rightarrow \Xi^{-}e^{+}\nu_{e}}$; $\rm{\Lambda_{c}^{+} \rightarrow \Lambda e^{+}\nu_{e}}$
\end{enumerate}
The Silicon-Pixel Detector (SPD), which constitutes the two innermost layers of the Inner Tracking System (ITS), is used as a multiplicity estimator via the number of track segments reconstructed from two hits in the two layers of the SPD, pointing to the primary vertex. In addition, multiplicity is estimated based on the percentile distribution of the V0 amplitude in the V0 detector. The multiplicity intervals are converted to charged-particle multiplicity density (\dnchdeta) intervals within pseudorapidity $|\eta| < 1$. The ITS is also used for secondary-vertex reconstruction and tracking. The particle identification is carried out by the Time Projection Chamber (TPC), the Time Of Flight (TOF) detector, and the Electromagnetic Calorimeter (EMCal) for high \pt electrons. The muons are reconstructed at forward rapidity (2.5 <$ y $< 4) using the muon spectrometer. 
\section{Results}

\subsection{Heavy-flavor hadron production in pp collisions}

\begin{figure}[!ht]
    \centering
    \includegraphics[scale = 0.4]{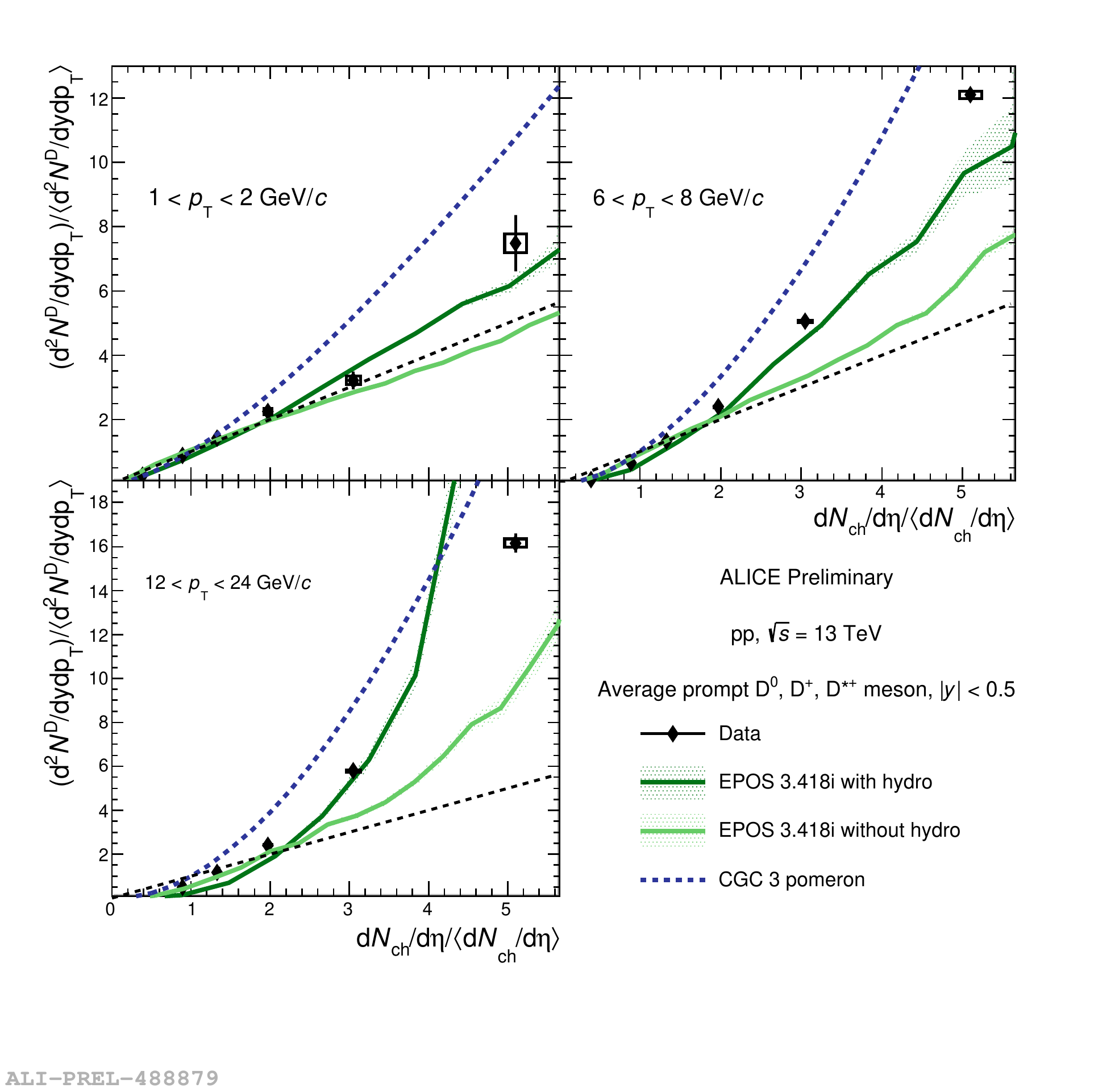}
\caption{\footnotesize{Average $\rm{D^0}$, $\rm{D^+}$, $\rm{D^{*+}}$ meson self-normalized yields in various \pt intervals as a function of charged-particle multiplicity in pp collisions at \sqs = 13 TeV compared to different model predictions\cite{Werner:2013tya,Schmidt:2020fgn}. The systematic uncertainties on D-meson yield as well as on \dnchdeta/$\langle$\dnchdeta$\rangle$} is depicted as boxes.}
    \label{fig:DmesonSNY}
\end{figure}

Heavy-flavor hadron production measurements as a function of multiplicity are expressed in the form of self-normalized yields, where the yields in the corresponding multiplicity interval ($\rm{Y^{mult}}$) are presented relative to those in the integrated multiplicity sample ($\rm{Y^{mult\ int}}$).
\begin{equation}
\rm{Y_{corr}^{mult}} = \left(\frac{Y^{mult}}{((Acc\times \epsilon_{prompt}^{mult})\times \it{N}\rm_{event}^{mult})/\epsilon^{trg}_{mult}}\right) \Biggm/ \left(\frac{Y^{mult\ int}}{((Acc \times \epsilon_{prompt}^{mult\ int})\times \it{N}\rm_{event}^{mult\ int})/\epsilon^{trg}_{mult\ int}}\right)
\label{selfnormalizedyield}
\end{equation}
Here, $\rm{Y^{mult}}$ is the raw yield, $\rm{Acc\times \epsilon_{prompt}^{mult}}$ is the acceptance$\times$efficiency factor, $\rm{\it{N}\rm^{mult}_{event}}$ is the number of events, and $\rm{\epsilon^{trg}_{mult}}$ is the trigger efficiency for a particular multiplicity interval. In Fig.~\ref{fig:DmesonSNY}, the average D-meson self-normalized yields at midrapidity ($|y|<0.5$) as a function of charged-particle multiplicity in pp collisions at \sqs = 13 TeV are presented for different \pt intervals. The multiplicity dependence of D-meson yields shows a faster than linear increase. They are compared to model predictions from EPOS3~\cite{Werner:2013tya} and the 3-pomeron Color Glass Condensate model (CGC)~\cite{Schmidt:2020fgn}. EPOS3 with a hydrodynamic evolution shows good agreement with a faster than linear increase of D-meson yields at low and intermediate \pt. EPOS3 without hydrodynamics predicts a small increase, leading to underestimation of the results. The 3-pomeron CGC largely overestimates the self-normalized yields of D-mesons. 
\begin{figure}[!ht]
    \flushleft
    \includegraphics[scale = 0.35]{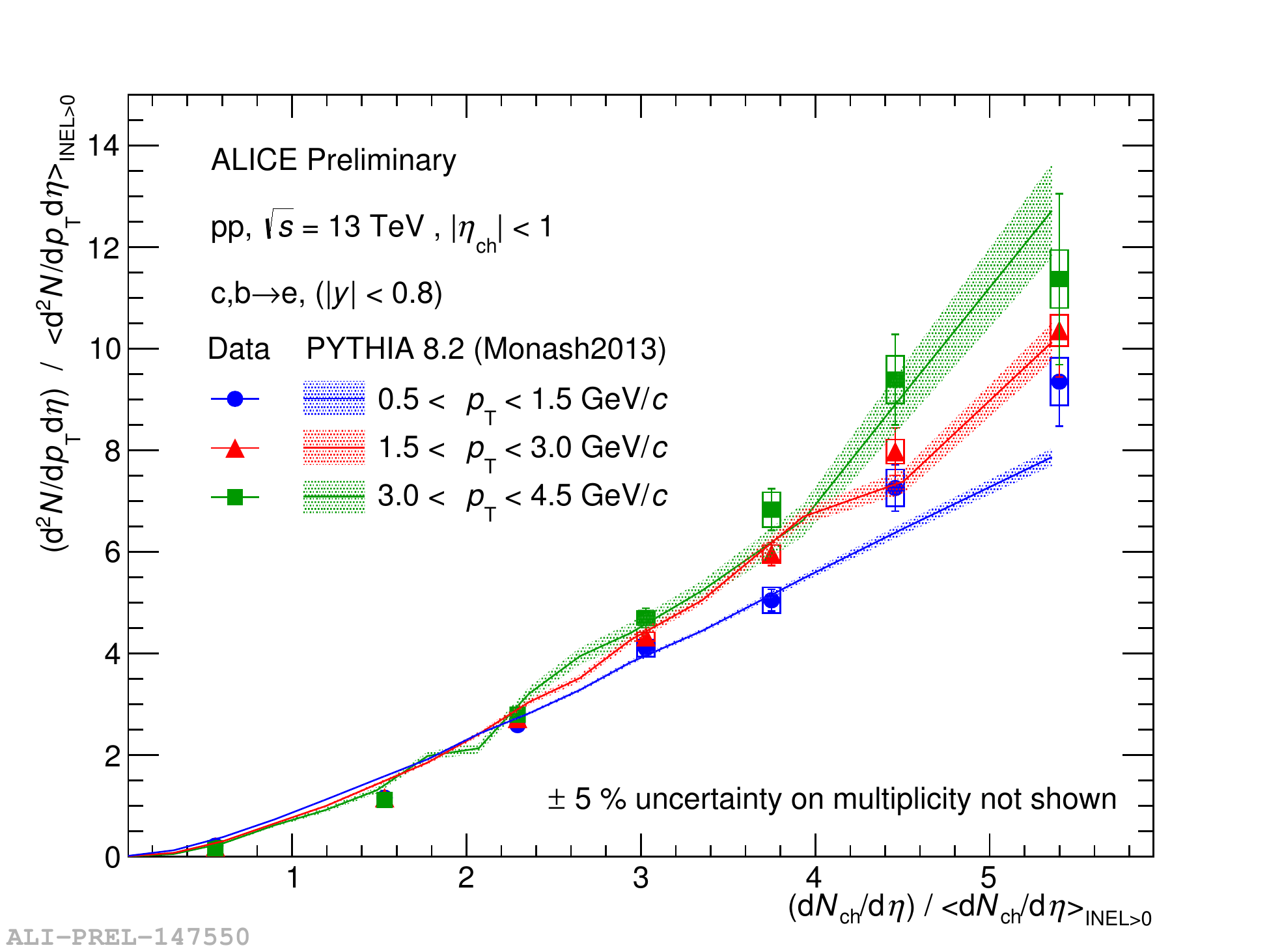}
    \includegraphics[scale = 0.39]{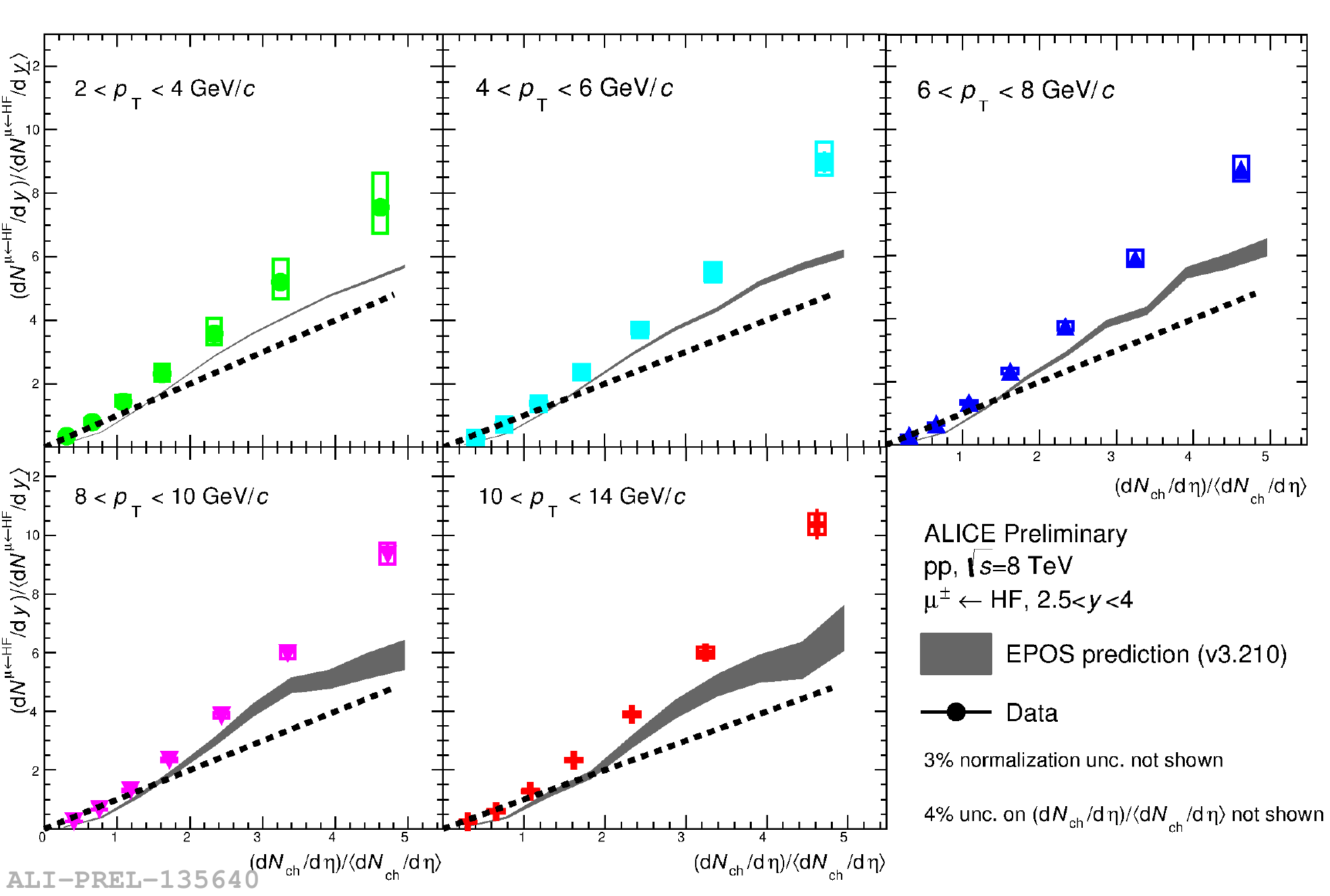}
    \caption{\footnotesize{Self-normalized yields of heavy-flavor hadron decay electrons measured in pp collisions at \sqs = 13 TeV (left) and muons measured in pp collisions at \sqs = 8 TeV (right) as a function of charged-particle multiplicity. The measurements are compared to model predictions from PYTHIA8 and EPOS3\cite{Sjostrand:2014zea,Werner:2013tya}.}}
    \label{fig:leptonSNY}
\end{figure}

\begin{figure}[!ht]
    \centering
    \includegraphics[scale = 0.3]{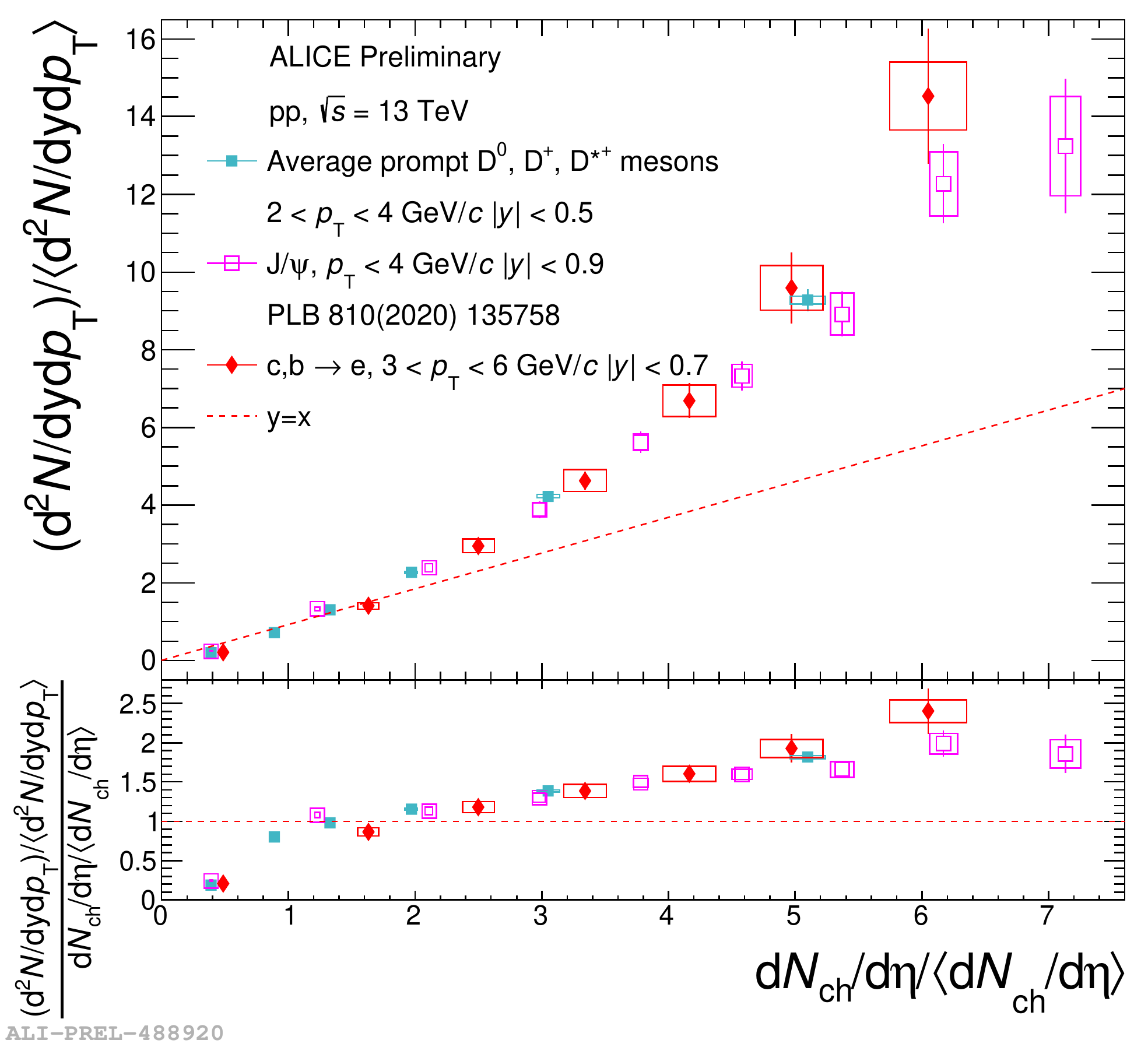}
    \includegraphics[scale = 0.3]{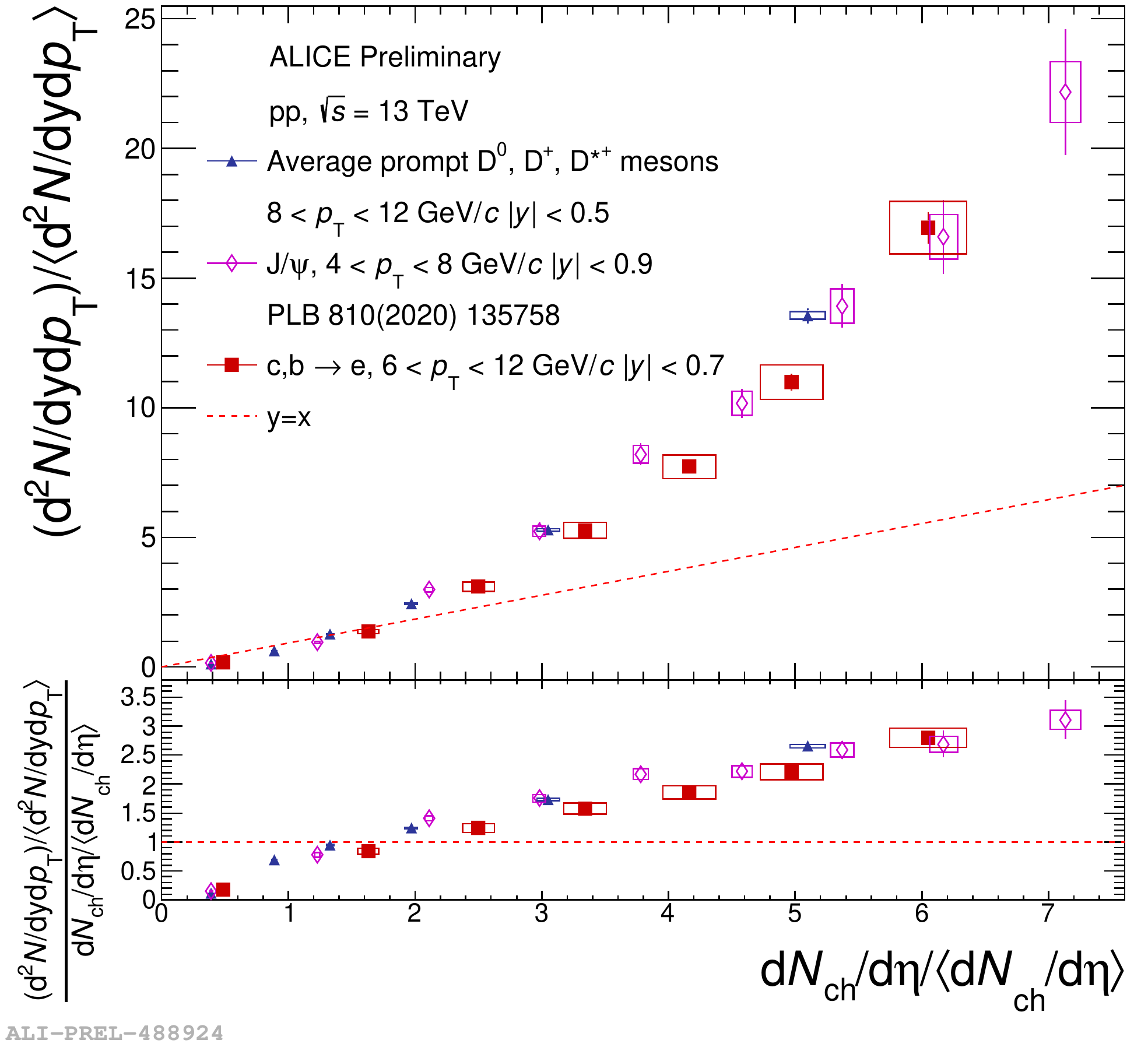}
    \caption{\footnotesize{Average D-meson, J/$\psi$~\cite{ALICE:2020msa}, and heavy-flavor hadron decay electrons self-normalized yields as a function of charged-particle multiplicity at midrapidity in pp collisions at \sqs = 13 TeV at compatible low (left) and high (right) \pt intervals. The bottom panels show the comparison of double ratios between the species.}}
    \label{fig:comparison}
\end{figure}

The self-normalized yields of electrons from heavy-flavor hadron decays in pp collisions at midrapidity at \sqs = 13 TeV for \pt < 4.5 GeV/$c$ are shown in the left panel of Fig.~\ref{fig:leptonSNY}. The results are compared to PYTHIA8 Monash tune with MPI and CR turned on\cite{Sjostrand:2014zea}. Possible auto-correlation effects that could arise from the measurement of the charged-particle distribution in the same pseudorapidity region as the charm hadrons could play a role, producing the faster than linear trend as a function of charged-particle multiplicity~\cite{Weber:2018ddv}. Muon production from the decay of heavy-flavor hadrons is measured at forward rapidity in pp collisions at \sqs = 8 TeV, shown on the right panel of Fig.~\ref{fig:leptonSNY}. The results are underestimated by predictions from EPOS3. In Fig.~\ref{fig:comparison}, the same results are put together with self-normalized yields from J/$\psi$\cite{ALICE:2020msa} and electrons from heavy-flavor hadron decays as a function of charged-particle multiplicity in pp collisions at \sqs = 13 TeV at compatible \pt intervals.

\begin{figure}[!ht]
    \centering
    \includegraphics[scale = 0.3]{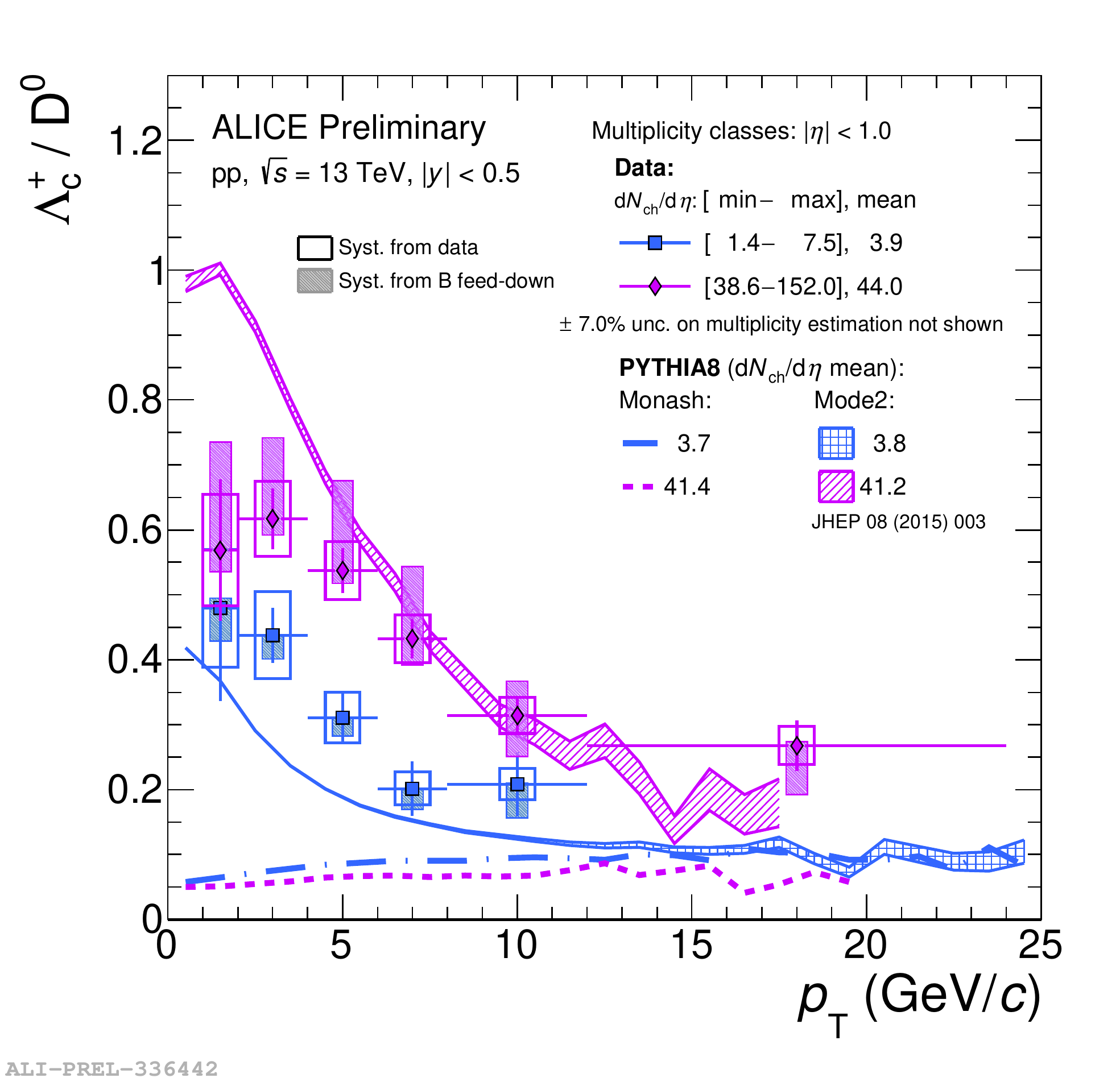}
    \includegraphics[scale = 0.3]{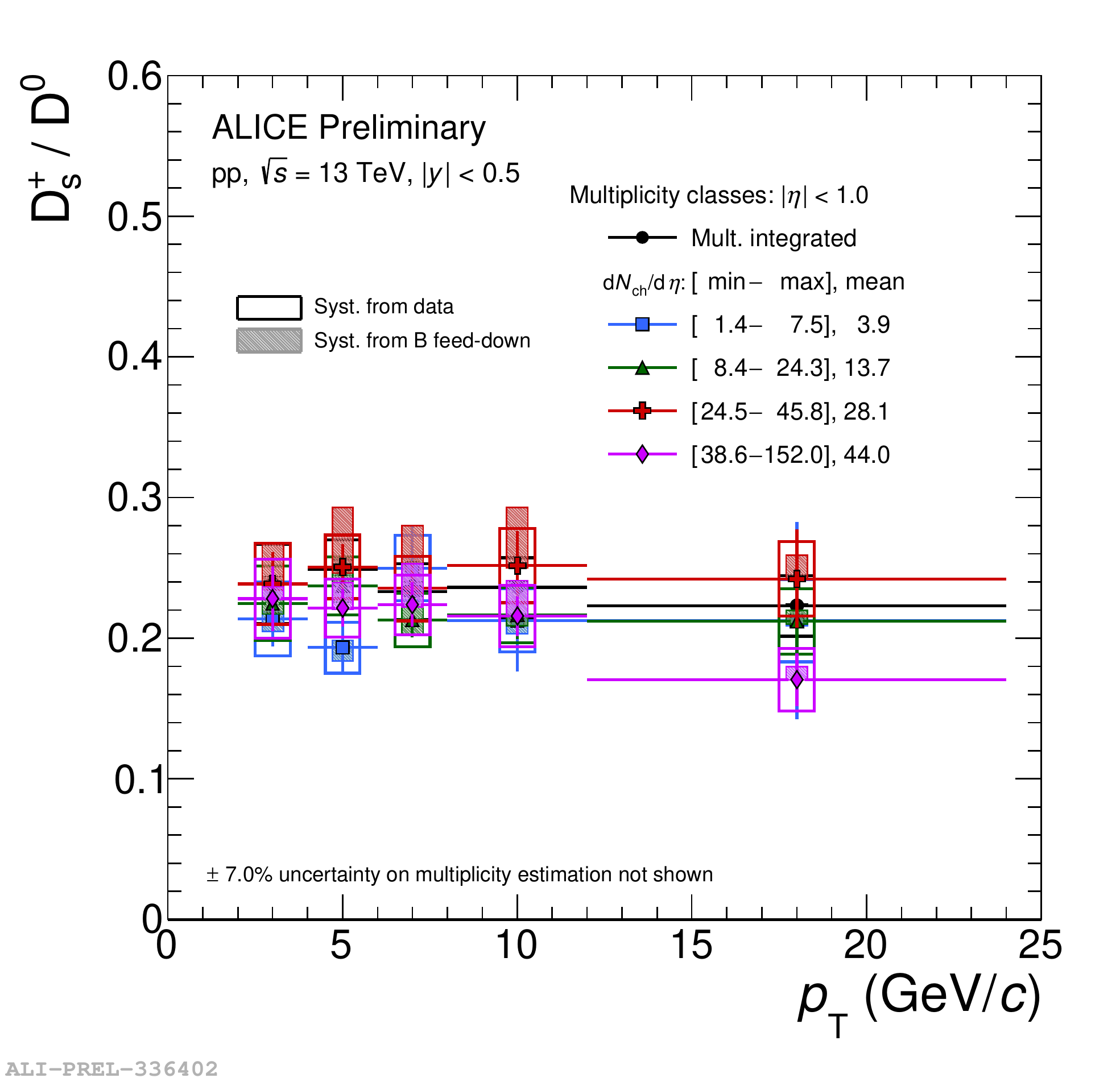}
    \caption{\footnotesize{(Left) $\rm{\Lambda_{c}^{+}/D^{0}}$ and (right) $\rm{D_{s}^{+}/D^{0}}$ yield ratios in different multiplicity intervals versus \pt measured in pp collisions at \sqs = 13 TeV. Comparisons are shown for $\rm{\Lambda_{c}^{+}/D^{0}}$ ratios with different PYTHIA8 tunes~\cite{Sjostrand:2014zea,Christiansen:2015yqa}.}}
    \label{fig:LcDmult}
\end{figure}

 The $\rm{\Lambda_{c}^{+}/D^{0}}$ and $\rm{D_{s}^{+}/D^{0}}$ yield ratios in pp collisions at \sqs = 13 TeV as a function of \pt, in different multiplicity intervals are presented in Fig.~\ref{fig:LcDmult}. The $\rm{\Lambda_{c}^{+}/D^{0}}$ ratio depicts a multiplicity dependence that hints to a modification of hadronization mechanisms with multiplicity. The PYTHIA8 (Monash)~\cite{Sjostrand:2014zea} tuned to $\rm{e^+e^-}$ collisions highly underestimates these baryon-to-meson yield ratios. The trend with multiplicity is qualitatively described by PYTHIA8 with CR Mode 2~\cite{Christiansen:2015yqa}. The $\rm{D_{s}^{+}/D^{0}}$ ratios do not show any visible multiplicity dependence, and the results are compatible with the \pt-integrated measurements at $\rm{e^+e^-}$ collisions~\cite{Gladilin:2014tba}.
\subsection{Heavy-flavor hadron production in p--Pb collisions}

\begin{figure}[!ht]
    \centering
    \includegraphics[scale = 0.27]{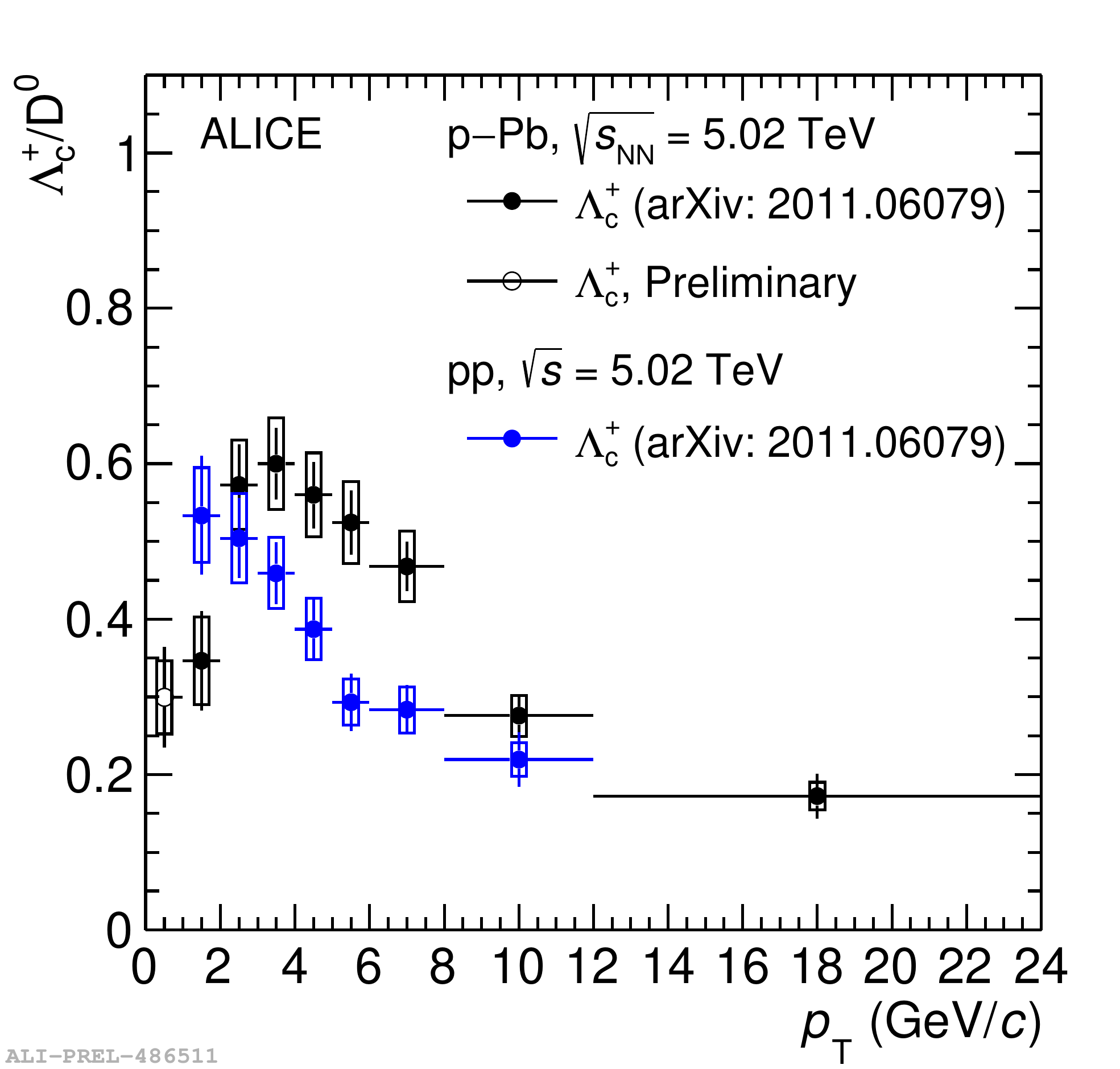}
    \includegraphics[scale = 0.47]{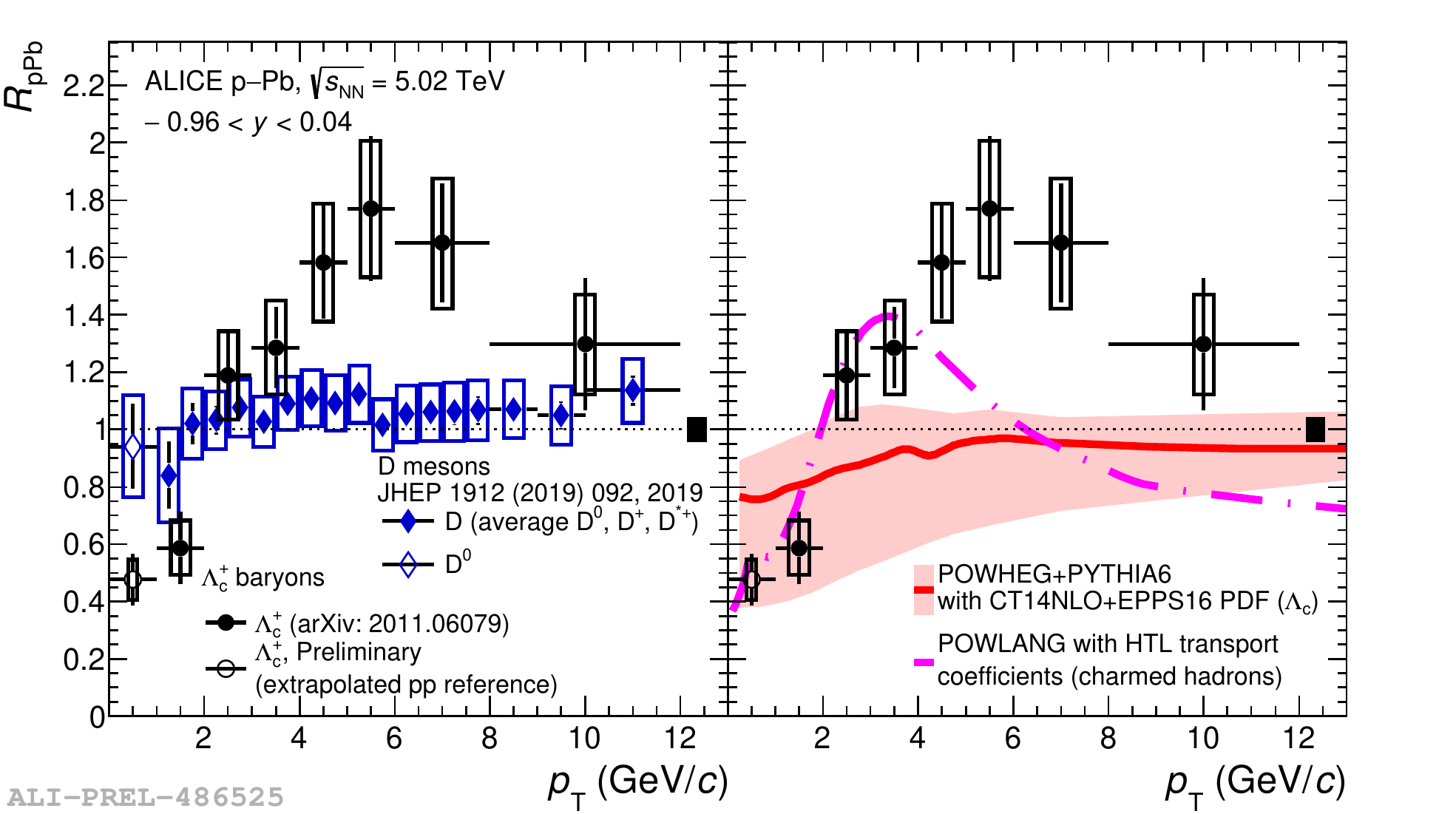}
    \caption{\footnotesize{(Left) $\rm{\Lambda_{c}^{+}/D^{0}}$ as as function of \pt measured in p--Pb collisions at \sqsn = 5.02 TeV~\cite{ALICE:2020wla}. (Right) $\rm{\Lambda_{c}^{+}}$ $R_{\rm pPb}$ as a function of \pt compared to D-meson results~\cite{ALICE:2019fhe} and model predictions\cite{Frixione_2007,Sj_strand_2006,Beraudo:2015wsd}.}}
    \label{fig:pPb}
\end{figure}

\begin{figure}[!ht]
    \centering
    \includegraphics[scale = 0.4]{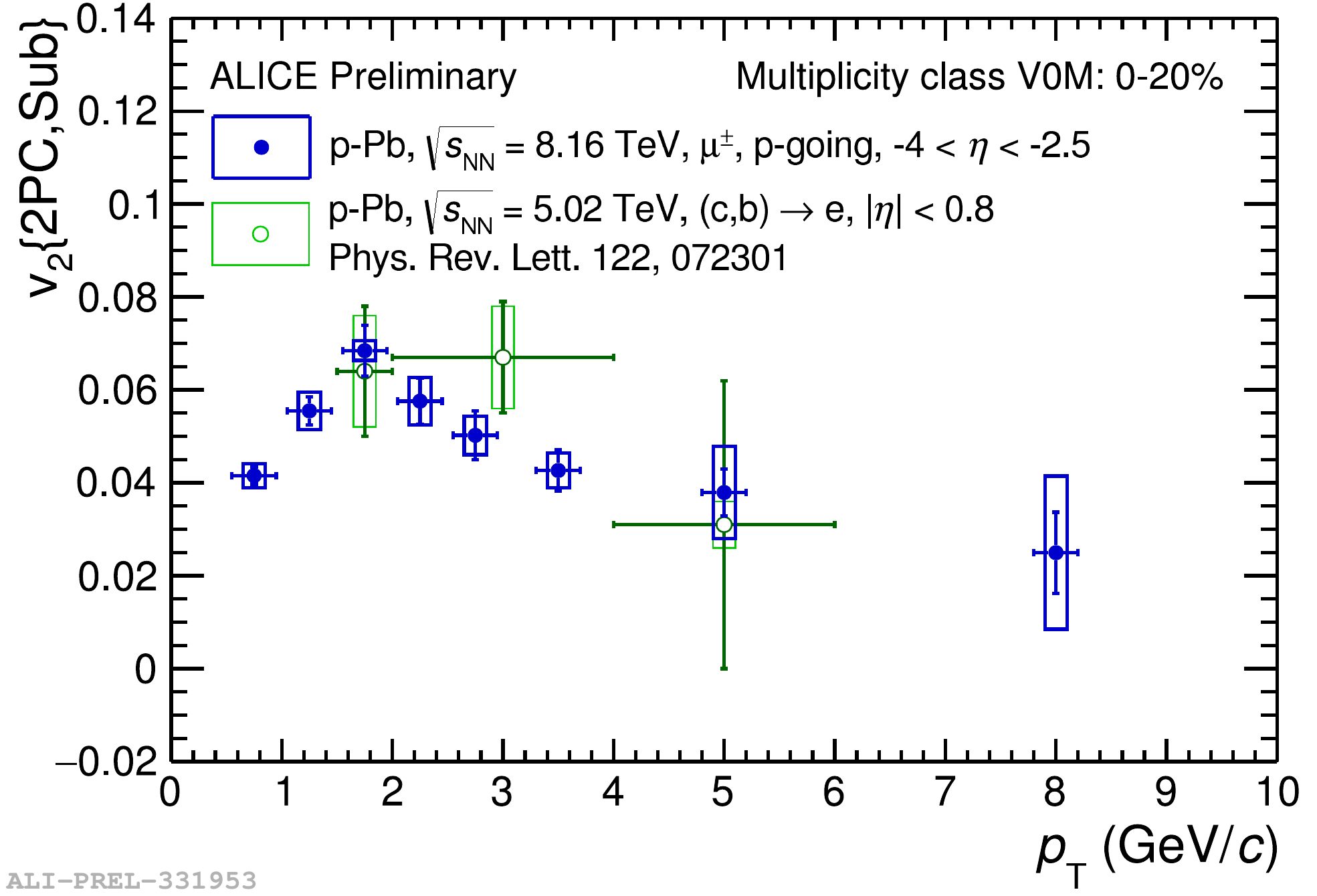}
    \caption{\footnotesize{Elliptic flow of inclusive muons as a function of \pt in p--Pb collisions at \sqsn = 8.16 TeV compared to the results for heavy-flavor hadron decay electrons at \sqsn = 5.02 TeV~\cite{ALICE:2018gyx}.}}
    \label{fig:HFmu}
\end{figure}

In Fig.~\ref{fig:pPb}, the $\rm{\Lambda_{c}^{+}/D^{0}}$ ratios measured in p--Pb collisions at \sqsn = 5.02 TeV is presented~\cite{ALICE:2020wla}. The most recent results extend the measurement down to \pt = 0 GeV/$c$. At low \pt, the $\rm{\Lambda_{c}^{+}/D^{0}}$ ratios in p--Pb collisions is lower than in pp collisions, the opposite is observed at intermediate \pt. This can be a consequence of radial flow or a modification in the hadronization mechanism in p--Pb systems. The nuclear modification factor which is defined as:
\begin{equation}
    R_{\rm{pPb}} = \frac{\rm{d\sigma_{pPb}/d\it{p}_{\rm{T}}}}{A\cdot\rm{d\sigma_{pp}/d\it{p}_{\rm{T}}}}
    \label{RAA}
\end{equation}
helps to quantify the CNM effects. In (\ref{RAA}) $\rm{d\sigma_{pp(pPb)}/d\it{p}_{\rm{T}}}$ are the \pt-differential cross sections in pp and p--Pb collisions for a corresponding center of mass energy, and $A$ is the atomic mass number of Pb. In Fig.~\ref{fig:pPb}, the $R_{\rm pPb}$ results are shown for p--Pb collisions at \sqsn = 5.02 TeV. In the left panel, the $\rm{\Lambda_{c}^{+}}$ $R_{\rm pPb}$ is compared to D-meson $R_{\rm pPb}$~\cite{ALICE:2019fhe}, in which a clear suppression is seen at low \pt for $\rm{\Lambda_{c}^{+}}$ followed by an enhancement at intermediate \pt. In the right panel, model calculations by POWHEG + PYTHIA6~\cite{Frixione_2007,Sj_strand_2006} and the POWLANG transport model~\cite{Beraudo:2015wsd} are shown. POWHEG with PYTHIA6 assumes hadronization via fragmentation, and the model agrees with the results at low \pt, however departing at intermediate \pt. The POWLANG model assumes the formation of a hot deconfined medium, implementing hadronization via fragmentation and quark recombination. The model predicts the suppression for \pt < 3 GeV/$c$, as it incorporates nuclear shadowing, although it deviates from the measurements at intermediate and high \pt.
In Fig.~\ref{fig:HFmu}, the elliptic flow ($v_2$) of heavy-flavor hadron decay leptons in p--Pb collisions is displayed. In general, the $n^{th}$-order anisotropic flow coefficients are written as $v_{n} = \langle \cos[n(\phi-\Psi_{R})]\rangle$, where $\phi$ is the azimuthal angle of a particle and $\Psi_{R}$ is the $n^{th}$ harmonic symmetry plane angle. $v_2$ is sensitive to the initial collision geometry and has dominant contribution in non-central collisions\cite{Poskanzer:1998yz}. The measurement for inclusive muons was performed at 2.5 <$ y $< 4 at \sqsn = 8.16 TeV and for electrons at $|y|<0.8$ at \sqsn = 5.02 TeV~\cite{ALICE:2018gyx}. The muon production is dominated by heavy-flavor hadron decays for \pt > 2 GeV/$c$. The results show a positive $v_2$ in central p--Pb events, which hints at the possibility of collective phenomena in high-multiplicity p--Pb collisions.

\section{Conclusion}

The self-normalized yield measurements of heavy-flavor hadron production as a function of multiplicity show a stronger than linear trend. Measurements of D-meson, J/$\psi$, and heavy-flavor decay leptons are compatible in similar \pt intervals. These measurements are compared to different theoretical calculations. In p--Pb collisions, measurements of $\rm{\Lambda_{c}^{+}/\rm{D^0}}$ yield ratios down to \pt = 0 GeV/$c$ at \sqsn = 5.02 TeV indicate a deviation from measurements in pp collisions. The $\rm{\Lambda_{c}^{+}}$ $R_{\rm pPb}$ shows clear differences at low and intermediate \pt when compared to D-meson measurements. Finally, a positive $v_2$ is measured for inclusive heavy-flavor decay muons in central p--Pb collisions at 2.5 <$ y $< 4 and heavy-flavor decay electrons at $|y|<0.8$.
\bibliography{SciPost_Example_BiBTeX_File.bib}





\nolinenumbers

\end{document}